\documentstyle[amssymb]{fbssuppl}

\input psfig.sty

\begin{document}
\title{The Few-Body Coulombian Problem}

\author{E.O. Alt\thanks{{\it E-mail address:} alt@dipmza.physik.uni-mainz.de}}
\institute{Institut f\"ur Physik, Universit\"at Mainz, D-55099 Mainz, Germany }
\date{}
\maketitle

\begin{abstract}
Recent advances in the treatment of scattering of charged composite particles are reviewed. In a first part I report on developments of the theory. Specifically I describe the recent completion of the derivation of the co-ordinate space asymptotic behaviour of the wave function for three charged particles in the continuum. This knowledge is increasingly being made use of in attempts to `derive' three-Coulomb particle wave functions to be used in all of configuration space which are solutions of the Schr\"odinger equation, though not everywhere but at least in one or preferably all of the asymptotic regions. Their practical application in approximate calculations of ionisation and breakup processes is pointed out. The asymptotic three-charged particle wave functions find further use in investigations of asymptotic and analytic properties of matrix elements of the three-body Coulomb resolvent. An important example is the {\it nonperturbative} derivation, valid for all energies, of the large-distance behaviour of the optical potential. In the second part I describe a renewed attempt to establish the few-body approach as a valuable tool for calculating (energetic) atomic collision processes. At the end I briefly touch upon the recent successful above-breakup-threshold calculation of proton-deuteron elastic scattering for a realistic potential. 
\end{abstract}

\section{Theoretical Developments}

\subsection{Asymptotic Wave Function for Three Charged Particles in the Continuum}

\subsubsection{Introduction} 
Knowledge of the asymptotic form of the wave function is required, e.g., as boundary condition for selecting the physical solution of the Schr\"odinger equation (SE) for energies above the ionisation threshold, and for deriving asymptotic and/or analytic properties of quantities which are matrix elements of the three-body Coulomb resolvent, such as the optical potential or the kernel of momentum space integral equations for charged particles (for a discussion of the latter topic, see \cite{maa98}). And, last but not least, it avails also for approximate calculations of breakup or ionisation in peripheral collisions.

I start by introducing some notation. Consider three distinguishable particles with masses $m_{\nu}$ and charges $e_{\nu},\,\nu = \alpha, \beta, \gamma$. In the c.m.\ system Jacobi co-ordinates are used: ${\vec k}_{\alpha} \,({\vec r}_{\alpha})$ is the relative momentum (co-ordinate) between particles $\beta$ and $\gamma$, and $\vec q_{\alpha} \,({\vec \rho}_{\alpha})$ the momentum (co-ordinate) of particle $\alpha$ relative to the center of mass of the pair $(\beta, \gamma)$. Frequently, two Jacobi three-vectors are combined into a sixdimensional vector: $\vec P:=({\vec k}_{\nu}, \,\vec q_{\nu})$ and $\vec X:=({\vec r}_{\nu}, \,{\vec \rho}_{\nu}), \,\nu = \alpha, \beta, \gamma$.

For three particles in the continuum the energy shell condition reads $E = q_{\alpha}^{2}/2 M_{\alpha} + k_{\alpha}^{2}/2 \mu_{\alpha}$, with the standard definitions $\mu_{\alpha}^{-1} = m_{\beta}^{-1} + m_{\gamma}^{-1}$ and $M_{\alpha}^{-1} = m_{\alpha}^{-1} + (m_{\beta}+m_{\gamma})^{-1} $ of the appropriate reduced masses. The interaction between particles $\beta$ and $\gamma$ is assumed to consist of a short-range and a Coulomb part
\begin{eqnarray}
V_{\alpha} = V_{\alpha}^S + V_{\alpha}^C, \quad \mbox{\rm with}\quad V_{\alpha}^C(r_{\alpha}) = \frac{e_{\beta} e_{\gamma}}{r_{\alpha}}. \label{pot}
\end{eqnarray} 
For the following it proves convenient to eliminate the three-body plane from the wave function by introducing a `reduced' wave function $\tilde \Psi^{(+)}_{ \vec P}(\vec X) $ via
\begin{eqnarray} 
\Psi^{(+)}_{ \vec P}(\vec X) = :\; e^{ \i \vec P \cdot \vec X }\; \tilde \Psi^{(+)}_{ \vec P}(\vec X) \equiv e^{ \i ({\vec k}_{\alpha} \cdot {\vec r}_{\alpha} + \vec q_{\alpha} \cdot {\vec \rho}_{\alpha})} \, \tilde {\Psi}^{(+)}_{ {\vec k}_{\alpha} {\vec q}_{\alpha}}({\vec r}_{\alpha} ,{\vec \rho}_{\alpha}) . \label{rwfn}
\end{eqnarray}

\subsubsection{Asymptotic Solution of the SE in $\Omega _{0}$: `Redmond Asymptotics'.}

This region of configuration space is reached by letting $r_{\alpha}, r_{\beta},$ and $r_{\gamma}$ approach infinity, but such that {\em not } $r_{\nu}/\rho_{\nu} \rightarrow 0 \;\mbox{for }\, \nu = \alpha, \beta, \mbox{ or } \gamma$; in other words, asymptotically all inter-particle distances grow uniformly. Introduce the Coulomb parameter, e.g., for the subsystem $(\beta, \gamma)$ as $\eta_{\alpha} = e_{\beta} e_{\gamma}\mu_{\alpha}/ k_{\alpha}$. Then, as stated by Redmond \cite{r72}, in leading order one has
\begin{eqnarray} 
\tilde \Psi^{(+)}_{ \vec P}(\vec X) \; \stackrel{\Omega _{0}}{\approx} \; \prod_{\nu = 1}^3 e^{ \i \eta_{\nu} \ln( k_{\nu} r_{\nu}- {\vec k}_{\nu} \cdot {\vec r}_{\nu})} \label{psi0}
\end{eqnarray} 
in the so-called `non-singular regions' (i.e., for $k_{\nu} r_{\nu} \neq {\vec k}_{\nu} \cdot {\vec r}_{\nu} $). An asymptotically equivalent form is 
\begin{eqnarray} 
\tilde \Psi^{(+)}_{ \vec P}(\vec X) \; \stackrel{\Omega _{0}}{\approx} \; \prod_{\nu = 1}^3 {\tilde {\psi}}_{ {\vec k}_{\nu }}({\vec r}_{\nu }), \label{psi0'}
\end{eqnarray} 
where ${\tilde {\psi}}_{{\vec k}_{\alpha}}({\vec r}_{\alpha})$, when multiplied with the appropriate plane wave, is the continuum solution of two-body SE 
\begin{eqnarray} 
\left\{ \frac {k_{\alpha}^{2}}{2 \mu_{\alpha}} + \frac{\Laplace_{{\vec r}_{\alpha}}}{2 \mu_{\alpha}} - V_{\alpha}^C({\vec r}_{\alpha}) - V_{\alpha}^S({\vec r}_{\alpha}) \right\} e^{ \i {\vec k}_{\alpha} \cdot {\vec r}_{\alpha}} {\tilde {\psi}}_{ {\vec k}_{\alpha}}^{(+)}({\vec r}_{\alpha}) = 0, \label{se2}
\end{eqnarray}
with leading asymptotic behaviour 
\begin{eqnarray}
{\tilde {\psi}}_{ {\vec k}_{\alpha}}^{(+)}({\vec r}_{\alpha}) \stackrel{r_{\alpha} \to \infty}{\longrightarrow } \; e^{ \i \eta_{\alpha} \ln( k_{\alpha} r_{\alpha}- {\vec k}_{\alpha} \cdot {\vec r}_{\alpha})}. \label{psi2as}
\end{eqnarray}
For purely Coulombic interactions this is just the known asymptotic form of the explicit solution of (\ref{se2}),
\begin{eqnarray} 
{\tilde {\psi}}_{ {\vec k}_{\alpha}}^{(+)}({\vec r}_{\alpha}) \equiv {\tilde {\psi}}_{ {\vec k}_{\alpha}}^{C(+)}({\vec r}_{\alpha}) = N _{\alpha} F(- \i \eta _{\alpha}, 1; \i ( k_{\alpha} r_{\alpha} - {\vec k}_{\alpha} \cdot {\vec r}_{\alpha})) \quad (V_{\alpha}^S \equiv 0 ).\label{psi2c}
\end{eqnarray} 
Here, $N_{\alpha} = e^{- \pi \eta_{\alpha}/2}\, \Gamma (1 + \i \eta_{\alpha})$; $F(a,b;x)$ is the confluent hypergeometric, and $\Gamma (z)$ the Gamma function. In the general case the behaviour (\ref{psi2as}) follows from the fact that for sufficiently large $r_{\alpha}$, $V_{\alpha}^S({\vec r}_{\alpha}) $ can be neglected in (\ref{se2}) yielding ${\tilde {\psi}}_{ {\vec k}_{\alpha}}^{(+)}({\vec r}_{\alpha}) \stackrel{r_{\alpha} \to \infty}{\longrightarrow } {\tilde {\psi}}_{ {\vec k}_{\alpha}}^{C(+)}({\vec r}_{\alpha})$. I, finally, mention that the form (\ref{psi0'}) with purely Coulombic wave functions (\ref{psi2c}) has been proposed in \cite{m77,bbk89}.

\subsubsection{Asymptotic Solution of SE in $\Omega _{\nu}, \;\nu = \alpha,\beta \mbox{ or } \gamma$: `AM Asymptotics'.}  \label{3corr}

Besides $\Omega _{0}$, there exist three more asymptotic regions in which the boundary condition on the wave function must be prescribed. They are characterised by
\begin{eqnarray}
\Omega _{\nu }: \quad \rho_{\nu } \rightarrow \infty ,\; r_{\nu }/\rho_{\nu } \rightarrow 0, \quad \mbox{for}\quad \nu = \alpha,\beta, \mbox{ or } \gamma.
\end{eqnarray}
That is, e.g. in $\Omega _{\alpha}$, the distance $\rho_{\alpha}$ of particle $\alpha$ from the center of mass of the pair $(\beta, \gamma)$ grows faster than the relative separation $r_{\alpha}$ between the particles of this pair. Recall the asymptotic behaviour (\ref{psi0}) in $\Omega _{0}$ which we rewrite as
\begin{eqnarray} 
\tilde \Psi^{(+)}_{ \vec P}(\vec X) \; \stackrel{\Omega _{0}}{\approx} \; e^{ \i \eta_{\alpha} \ln( k_{\alpha} r_{\alpha}- {\vec k}_{\alpha} \cdot {\vec r}_{\alpha})} \,\prod_{\nu \neq \alpha} e^{ \i \eta_{\nu} \ln( k_{\nu} r_{\nu}- {\vec k}_{\nu} \cdot {\vec r}_{\nu})} . \label{psi0''}
\end{eqnarray} 
This suggests that to go from $\Omega_0$ to ${\Omega _{\alpha}}$, which includes all values $r_{\alpha} \in [0,\infty)$ while still $r_{\beta}, \, r_{\gamma} \to \infty$, it might suffice to replace in (\ref{psi0''}) the asymptotic wave function (\ref{psi2as}) for the pair $(\beta, \gamma)$ by the full solution ${\tilde {\psi}}_{{\vec k}_{\alpha}}({\vec r}_{\alpha})$ of the SE (\ref{se2}).

However, this simple idea turns out to be not correct. Instead, as has been shown by Alt and Mukhamedzhanov \cite{am92} (AM) the leading term is given by
\begin{eqnarray}
\tilde {\Psi}^{(+)}_{ {\vec k}_{\alpha} {\vec q}_{\alpha}}({\vec r}_{\alpha} ,{\vec \rho}_{\alpha}) \; \stackrel{\Omega _{\alpha}}{\approx} \; {\tilde {\psi}}^{(+)}_{ {\vec k}_{\alpha}({\vec \rho}_{\alpha})}({\vec r}_{\alpha})\,
\prod_{\nu \neq \alpha} e^{ \i \eta_{\nu} \ln( k_{\nu} r_{\nu}- {\vec k}_{\nu} \cdot {\vec r}_{\nu})}, \label{psia}
\end{eqnarray}
(for further and next-to-leading order contributions see \cite{ml96, ks96,kz97,mai98}). What appears is the exact continuum solution of two-body-like SE (for all ${\vec r}_{\alpha} \in \Omega _{\alpha}$)
\begin{eqnarray}
\left\{ \frac {k_{\alpha}^{2}({\vec \rho}_{\alpha})}
{2 \mu_{\alpha}} + \frac{\Laplace_{{\vec r}_{\alpha}}}{2 \mu_{\alpha}} -
V_{\alpha}^C({\vec r}_{\alpha}) - V_{\alpha}^S({\vec r}_{\alpha})\right\} 
e^{ \i {\vec k}_{\alpha}({\vec \rho}_{\alpha}) 
\cdot {\vec r}_{\alpha}} \; {\tilde {\psi}}^{(+)}_{ {\vec k}_{\alpha} 
({\vec \rho}_{\alpha}) }({\vec r}_{\alpha}) = 0. \label{se2l}
\end{eqnarray}
It describes the relative motion of $\beta$ and $\gamma$ at an energy which depends parametrically on the distance from particle $\alpha$, as determined by the {\em local} momentum
\begin{eqnarray} 
 {\vec k}_{\alpha}({\vec \rho}_{\alpha}) = {\vec k}_{\alpha} + 
\frac{\vec a_{\alpha}({\hat {\vec \rho}}_{\alpha})}{\rho_{\alpha}}, \quad \vec a_{\alpha}({\hat {\vec \rho}}_{\alpha}) = - \sum_{\nu \neq \nu' = \beta, \gamma} \eta_{\nu} \lambda_{\nu \nu'} \frac{\epsilon_{\alpha \nu} {\hat {\vec \rho}}_{\alpha} - {\hat { \vec k}}_{\nu}}{1 - \epsilon_{\alpha \nu} {\hat {\vec \rho}}_{\alpha} \cdot 
{\hat { \vec k}}_{\nu} } . \label{ploc}
\end{eqnarray}
Here, $ \lambda_{\nu \nu'} = m_{\nu}/(m_{\nu}+m_{\nu'}), \; {\hat { \vec k}}_{\nu}:= {\vec k}_{\nu}/{k}_{\nu}, \; \epsilon_{\alpha \beta} = - \epsilon_{\beta \alpha } = +1$ if $(\alpha,\beta)$ is a cyclic ordering of (1,2,3). If $V_{\alpha}^S \equiv 0$, the explicit solution of (\ref{se2l}) is
\begin{eqnarray} 
{\tilde {\psi}}_{ {\vec k}_{\alpha}({\vec \rho}_{\alpha})}^{C(+)} ({\vec r}_{\alpha}) = N_{\alpha}({\vec \rho}_{\alpha}) \; F(- \i \eta _{\alpha}({\vec \rho}_{\alpha}), 1; \i ( { k}_{\alpha}({\vec \rho}_{\alpha}) r_{\alpha}- {\vec k}_{\alpha}({\vec \rho}_{\alpha}) \cdot {\vec r}_{\alpha}) ,
\end{eqnarray} 
which is of the same form as (\ref{psi2c}) but with the asymptotic momentum ${\vec k}_{\alpha} $ replaced everywhere by the local momentum ${\vec k}_{\alpha}({\vec \rho}_{\alpha})$.

The physically decisive point is that the parametric dependence of the wave function ${\tilde {\psi}}^{(+)}_{ {\vec k}_{\alpha} ({\vec \rho}_{\alpha}) }$ for the pair $(\beta, \gamma)$ on the relative co-ordinate ${\vec \rho}_{\alpha}$ between its center of mass and particle $\alpha$ is a manifestation of a {\it long-ranged, non\-central, velocity-dependent, dynamic three-body correlation} which is a new genuine three-body effect. This is evident from rewriting the bracket in (\ref{se2l}) as 
\begin{eqnarray}
\bigg\{ \cdots \bigg\} = \left\{ \frac {k_{\alpha}^{2}}{2 \mu_{\alpha}} + \frac{\Laplace_{{\vec r}_{\alpha}}} {2 \mu_{\alpha}} - V_{\alpha}({\vec r}_{\alpha}) - \frac{1}{\rho_{\alpha}} \cdot \frac{\vec a_{\alpha}({\hat {\vec \rho}}_{\alpha}) \cdot { {\vec k}_{\alpha}}}{\mu_{\alpha}}\right\} + O \left(\rho_{\alpha}^{-2} \right). \label{psi2cl}
\end{eqnarray} 
Comparison with the bracket in the ordinary two-body SE (\ref{se2}) reveals that the additional, potential-like term in (\ref{psi2cl}) is indeed of long range ($\sim \rho_{\alpha}^{-1}$), and depends on the relative velocity ${\vec k}_{\alpha}/\mu_{\alpha}$ within the pair $(\beta, \gamma)$ and on the orientation of ${\vec k}_{\alpha}$ relative to ${\vec \rho}_{\alpha}$ (cf. the definition of ${\vec a}_{\alpha}$ in (\ref{ploc})). The latter properties can intuitively be understood as  a result of the `dynamic screening' of the charges when the three particles are in typical configurations pertaining to the asymptotic regions $\Omega_{\nu}, \, \nu = \alpha, \beta, \gamma$.

When can one expect to observe effects of these three-body correlations? As follows from (\ref{ploc}), local ($ {\vec k}_{\alpha}({\vec \rho}_{\alpha}) $) and asymptotic momentum ($ {\vec k}_{\alpha}$) will differ appreciably only if $ k_{\alpha}$ is small (recall that $\rho_{\alpha} \to \infty$ in $\Omega_{\alpha}$). As a measure of the `size' of their difference we can take $|{\vec a}_{\alpha}({\hat {\vec \rho}}_{\alpha})| $ which for $k_{\alpha} \to 0$ behaves as
\begin{eqnarray}
|{\vec a}_{\alpha}({\hat {\vec \rho}}_{\alpha})| \stackrel{ k_{\alpha} \to 0}{\; \sim \; } \frac{M_{\alpha}}{q_{\alpha}} \bigg| \left[ \frac{e_{\beta}}{m_{\beta}} - \frac{e_{\gamma}}{m_{\gamma}} \right] \bigg|. 
\end{eqnarray}
That is, it is inversely proportional to the relative velocity of particle $\alpha$ and the center of mass of the pair $(\beta, \gamma)$, and proportional to the famous charge-over-mass ratio difference of the pair $(\beta, \gamma)$ (`Post-acceleration effect').

\subsubsection{Asymptotically Correct Three-body Coulomb Wave Functions}

Recently many attempts at `deriving' three-body Coulomb wave functions, which extrapolate the known asymptotic wave functions into the non-asymptotic region, have appeared. Obviously, such are highly non-unique and, thus, open to further optimisation. For instance, a wave function which is asymptotically correct in all regions $\Omega _{1} \cup \Omega _{2} \cup \Omega _{3} \cup \Omega _{0}$ has been proposed in \cite{am92} 
\begin{eqnarray}
\tilde \Phi^{(+)}_{ \vec P}(\vec X) = \prod_ {\nu = 1}^{3} \tilde N _{\nu} F(- \i {\tilde \eta}_{\nu},1; \i ( {\tilde k}_{\nu} r_{\nu} - {\tilde {\vec k}}_{\nu} \cdot {\vec r}_{\nu})),
\end{eqnarray}
with ${\tilde {\vec k}}_{\alpha} = {\vec k}_{\alpha} + {2 {\vec a}_{\alpha}({\hat \rho}_{\alpha})}/R, \; R = \sum_{\nu = 1}^{3} r_{\nu},$ ${\tilde \eta}_{\nu}:=\eta _{\nu}({\tilde {\vec k}}_{\nu}), \; {\tilde N}_{\nu}:= N_{\nu}({\tilde \eta}_{\nu})$. A structurally identical wave function but with a different definition of the local momenta has recently been proposed in \cite{mmgcg97,cgg98}. For a wave function which contains - again differently defined - local momenta in only two hypergeometric functions and hence is asymptotically correct in three regions, say $\Omega _{\alpha} \cup \Omega _{\beta} \cup \Omega _{0}$, only see \cite{ekm97}. It is to be noted that all these admissible choices of local momenta differ from the original one \cite{am92} in next-to-leading order in $1/\rho_{\alpha}$ only.

Most effort has, however, been directed towards devising wave functions which are asymptotically correct in $\Omega_{0}$ only. The prototype is (\ref{psi0'}) (with ${\tilde {\psi}}_{ {\vec k}_{\alpha}}^{(+)} \to {\tilde {\psi}}_{ {\vec k}_{\alpha}}^{C(+)}$) which, though it has the unphysical feature that the motion of the three particle pairs is completely {\it uncorrelated in all space}, is nevertheless widely used. To overcome this deficiency correlations are introduced, by modifying the Coulomb parameter which occur in each of the hypergeometric functions (\ref{psi2c}): $\eta_{\alpha} \to \mu_{\alpha}[e_{\beta} e_{\gamma} + \chi_{\alpha}]/k_{\alpha}$ where either $\chi_{\alpha} = \chi_{\alpha}({\vec k}_{\alpha},{\vec q}_{\alpha})$ depends on the momenta (`velocity-dependent charges' \cite{jf92,bb94}) or on the positions $\chi_{\alpha} =\chi_{\alpha}({\vec r}_{\alpha}, {\vec \rho}_{\alpha})$ (`position-dependent charges' \cite{b96}). The fairly arbitrary functions $\chi_{\alpha}$ are chosen such that the resulting three-body wave functions satisfy additional criteria like the cusp condition. For very general wave functions of this type see \cite{cgg98}.

\subsubsection{Applications}

These - at least asymptotically correct - wave functions are increasingly being applied to approximate calculations of ionisation and breakup processes. In its simplest form (with incoming plane wave $|{\vec q}_{\alpha} \rangle$) the corresponding amplitude is 
\begin{eqnarray}
{\cal T}_{\alpha m,0}({\vec q}_{\alpha};{\vec k}_{\beta}',{\vec q}_{\beta}') = \langle \Psi^{(-)}_{{\vec k}_{\beta}',{\vec q}_{\beta}'}| \bar V_{\alpha} | \psi_{\alpha m}\rangle |{\vec q}_{\alpha} \rangle, \label{tbu}
\end{eqnarray}
with $\bar V_{\alpha} = \sum_{\nu \neq \alpha} V_{\nu} $ being the (entrance) channel interaction and $\psi_{\alpha m} = \psi_{\alpha m} ({\vec r}_{\alpha})$ denoting the incoming bound state wave function. Expression (\ref{tbu}) contains for the final state the full solution $\langle \Psi^{(-)}_{{\vec k}_{\beta}',{\vec q}_{\beta}'}|$ of the three-charged particle SE which is being approximated by one of the aforementioned ansaetze. 

Many calculations have appeared for (e,2e) reactions on hydrogen and other atoms; for a comparative study of various types of asymptotically correct wave functions see \cite{jmk97}. A nuclear physics example is the Coulomb dissociation of a projectile in the field of a fully stripped nucleus which is an important tool for obtaining astrophysical $S$-factors \cite{i98}. Quite generally one expects the better results with such an approximation the more peripheral the collision is (the impact parameter should be larger than the range of the nuclear interaction).

\subsection{Long-range Behaviour of the Optical Potential (OP)}

\subsubsection{Introduction} 
As is well known, elastic scattering of the type $\alpha + (\beta,\gamma)_m \to \alpha + (\beta,\gamma)_m$ can formally be described by means of a single, one-channel LS equation of the type
\begin{eqnarray}
{\cal T}_{\alpha m, \alpha m}(z) = {\cal V}_{\alpha m, \alpha m}^{\rm opt}(z) + {\cal V}_{\alpha m, \alpha m}^{\rm opt}(z) {\cal G}_{0;\alpha m}(z) {\cal T}_{\alpha m, \alpha m}(z) . \label{telst}
\end{eqnarray}
Though not useful for practical calculations, the very existence of an equation like (\ref{telst}) serves as justification for use of phenomenological OP's. 

The solvability of (\ref{telst}) depends on the properties of the plane wave matrix elements ${\cal V}_{\alpha m, \alpha m}^{\rm opt}({\vec q}_{\alpha}', \vec q_{\alpha};z) = \langle {\vec q}_{\alpha}'| {\cal V}_{\alpha m,\alpha m}^{\rm opt}(z) |{\vec q}_{\alpha}\rangle $ in the limit ${\vec \Delta}_{\alpha} = \vec q_{\alpha}'- \vec q_{\alpha} \to 0$ which reflects itself in a corresponding behaviour in co-ordinate space for large intercluster separation $\rho_{\alpha} $. The exact expression for the OP comes as a sum of two terms ${\cal V}_{\alpha m, \alpha m}^{\rm opt} \;=\; {\cal V}_{\alpha m, \alpha m}^{\rm stat} + {\tilde {\cal V}}_{\alpha m, \alpha m}^{\rm opt}$. The first term, so-called static potential, has only the trivial Coulomb-type behaviour which for a spherically symmetric target reads as ($e_{\beta \gamma}=e_{\beta}+e_{\gamma}$)
\begin{eqnarray}
{\cal V}_{\alpha m, \alpha m}^{\rm stat}({\vec q}_{\alpha}', {\vec q}_{\alpha}) \; \stackrel{\Delta_{\alpha} \to 0}{=} \; \frac{4\pi e_{\alpha}e_{\beta \gamma}} {\Delta_{\alpha}^{2}} \quad \Longleftrightarrow \quad 
{\cal V}_{\alpha m, \alpha m}^{\rm stat}({\vec \rho}_{\alpha}) \; \stackrel{\rho_{\alpha} \rightarrow \infty}{=} \; \frac{e_{\alpha}e_{\beta \gamma}}{ \rho_{\alpha}} . 
\end{eqnarray}

Thus it is the `nonstatic' part ${\tilde {\cal V}}_{\alpha m, \alpha m}^{\rm opt}$ which is of interest. For energies $E < 0$, i.e. below the three-body threshold, its behaviour is known from $2^{nd}$ order perturbation theory, and has been corroborated for the exact OP in \cite{km88}:
\begin{eqnarray} 
{\tilde {\cal V}}_{\alpha m, \alpha m}^{\rm opt}({\vec q}_{\alpha}', {\vec q}_{\alpha}; E) \, 
\stackrel{ \Delta_{\alpha}\rightarrow 0}{\sim} \, \Delta_{\alpha} \quad \Longleftrightarrow \quad 
{\tilde {\cal V}}_{\alpha m, \alpha m}^{\rm opt}({\vec \rho}_{\alpha}) \; \stackrel{\rho_{\alpha} \rightarrow \infty}{\approx}\; - \frac{a}{2 \rho_{\alpha}^4}.\label{as1}
\end{eqnarray}
Here, $a$ static dipole polarisability of the composite particle. 

But does this result hold also above the three-body threshold? In fact, the two {\it perturbative} answers are in support of it, but Ref. \cite{mss82} finds an energy-dependent $"a"$ while in \cite{uc89} the standard expression for $a$ is obtained. And what happens to the {\it exact} nonstatic OP for $E > 0$? Answering requires the investigation of the zero-momentum transfer limit of ${\tilde {\cal V}}_{\alpha m, \alpha m}^{\rm opt}(\vec q_{\alpha}',\vec q_{\alpha};z)$ which is fully determined by the Coulomb part of the channel interaction $\bar V_{\alpha}^C = \sum_{\nu \neq \alpha} V_{\nu}^C$ and by the three-body Coulomb resolvent $G^C(z) = (z-H_0 - \sum_{\nu} V_{\nu}^C)^{-1}$,
\begin{eqnarray}
{\tilde {\cal V}}_{\alpha m, \alpha m}^{\rm opt}(\vec q_{\alpha}',\vec q_{\alpha};z) \; &\stackrel{\Delta_{\alpha}\rightarrow 0}{\approx}& \; {\tilde {\cal V}}_{\alpha m, \alpha m}^{\rm opt \,(as)}(\vec q_{\alpha}',\vec q_{\alpha};z) \nonumber \\
&:=& \langle \vec q_{\alpha}'| \langle \psi_{\alpha m} | \bar V_{\alpha}^C Q_{\alpha m}G^C(z) Q_{\alpha m} \bar V_{\alpha}^C |\psi_{\alpha m} \rangle |\vec q_{\alpha} \rangle. \qquad
\end{eqnarray}
Here, $Q_{\alpha m}= 1 - |\psi_{\alpha m}\rangle \langle \psi_{\alpha m}|$ is the projector onto all (discrete and continuum) target states except the incoming (= outgoing) bound state.

The procedure goes as follows: insert the spectral representation of $ G^C$, with both two- and three-body scattering states; for the latter use the asymptotic three-charged particle wave function in $\Omega_{\alpha}$. Then extract the leading singularity in the limit $ \Delta_{\alpha}\rightarrow 0$. (Note: $2^{nd}$-order perturbation theory is equivalent to replacing $G^C$ by the Coulomb channel resolvent $G_{\alpha}^C = (z-H_0 - V_{\alpha}^C)^{-1}$.)

\subsubsection{Nonperturbative Result Valid for All Energies}

Let me just state the result without giving any details. For the latter I refer to \cite{am95}. Off the energy shell one finds
\begin{eqnarray} 
{\tilde {\cal V}}_{\alpha m, \alpha m}^{\rm opt\,(as)}({\vec q}_{\alpha}', {\vec q}_{\alpha}; E+i0) \; \stackrel{{\Delta}_{\alpha} \rightarrow 0}{=}\; C \Delta_{\alpha} + o\,(\Delta_{\alpha}) , \label{voptoff}
\end{eqnarray}
with an energy- and momentum-dependent `strength factor' $C=C(E, q_{\alpha})$. On the energy shell, i.e. for $E =q_{\alpha}^{2}/2 M_{\alpha} + {\hat E}_{\alpha m}, \, q_{\alpha}' = q_{\alpha}, \, ({\hat E}_{\alpha m}$ is the bound state energy), ${\tilde {\cal V}}_{\alpha m, \alpha m}^{\rm opt\,(as)}({\vec \Delta}_{\alpha})$ is a function of the momentum transfer only. Hence, in co-ordinate space one indeed ends up with a local, energy-independent potential
\begin{eqnarray} 
 {\tilde {\cal V}}_{\alpha m, \alpha m}^{\rm opt\,(as)}({\vec \rho}_{\alpha}) \; \stackrel{\rho_{\alpha} \to \infty}{=} \; 
- \frac{a}{2 \rho_{\alpha}^4} + o \left( \frac{1}{\rho_{\alpha}^4}\right). \label{coo}
\end{eqnarray}
A further important point is that the exact `strength factor' $a$ coincides with the polarisability as found in perturbative approaches,
\begin{eqnarray} 
a = 2 \sum_{n \neq m}\frac{|{\hat {\vec \rho}}_{\alpha} \cdot {\vec D}_{n m}|^2}{\left[ |{\hat E}_{\alpha m}| - |{\hat E}_{\alpha n}| \right]} + 2 \int \frac{d {\vec k}_{\alpha}^0}{(2\pi)^{3}}\frac{|{\hat {\vec \rho}}_{\alpha} \cdot {\vec D}_{{\vec k}_{\alpha}^0 m}|^2 }
{\left[ |{\hat E}_{\alpha m}| + {k_{\alpha}^0}^2/2 \mu_{\alpha} \right]},\\
\label{adip}
{\vec D}_{n m} =  \epsilon_{\alpha \beta} e_{\alpha}\mu_{\alpha} 
\left(e_{\gamma}/m_{\gamma} - e_{\beta}/m_{\beta}\right)
\int d {\vec r}_{\alpha} \psi_{\alpha n}^{*}({\vec r}_{\alpha}) 
{\vec r}_{\alpha} \psi_{\alpha m}
({\vec r}_{\alpha}) \\ \label{dnm3}
{\vec D}_{{\vec k}_{\alpha}^0 m}
 =  \epsilon_{\alpha \beta} e_{\alpha}\mu_{\alpha} 
\left(e_{\gamma}/m_{\gamma} - e_{\beta}/m_{\beta}\right)
\int d {\vec r}_{\alpha} \psi_{{\vec k}_{\alpha}^0}^{(+)*}
({\vec r}_{\alpha}) {\vec r}_{\alpha} 
\psi_{\alpha m}({\vec r}_{\alpha})
\label{dkm2}
\end{eqnarray} 
being the matrix elements of the dipole operator between the incoming and all other (bound and continuum) target states. That is, no "renormalisation" of $a$ arises from summing up all higher order terms in the perturbation expansion of $G^C$, and the $E$- and $q_{\alpha}$-dependence of the `off-shell strength' $C$ has disappeared.

Concluding I mention that a new nonrelativistic contribution to the asymptotic part of the OP $\sim 1/\rho_{\alpha}^5$, which results from the long-range three-body correlations described in Sect. (\ref{3corr}), has been derived in \cite{m97}.

\section{Calculational Developments}

Considerable progress has been made in recent years also in applying few-body theory to the practical computation of atomic and nuclear reactions. Here, I will concentrate on describing an attempt at re-establishing this method as a powerful tool for atomic physics calculations. Advances in the calculation of proton-deuteron scattering with realistic nuclear potentials will only briefly be touched upon (cf. the separate Contribution to this Conference \cite{ams98}).

\subsection{Few-Body Approach to the Energetic Atomic Collision ${\rm H}^+ + {\rm H}$} 

Scattering of protons off hydrogen has attracted the interest of physicists since the '30. Because direct scattering, electron exchange (both including excitation) and ionisation are coupled by unitarity they should in principle constitute an `ideal' field of application of three-body theory. However, except for a few isolated, and not so successful, attempts little has been accomplished in this respect. 

Instead, most efforts have concentrated on developing, and improving, standard models based on the SE. For instance, at low and medium energies the expansion into target states (`close coupling') appears to be the method of choice. At higher energies (from keV to a few hundreds of keV), which is the regime I will concentrate on, DWBA-type models or models based on the multiple scattering expansion of Lippmann-Schwinger equation for wave functions are rather successful. However, most of these approaches treat only one channel explicitly, i.e., intermediate-state rearrangement and/or excitation is ignored. One consequence is the violation of (three-body) unitarity constraints. Hence, though they generally provide a good description of either electron exchange or (in)elastic cross sections, they are frequently unable to describe the other reaction. For recent reviews I refer to \cite{de94,bmcd92}.

Two-fragment processes of the type $\alpha + (\beta,\gamma)_m \to \beta + (\alpha ,\gamma)_n$ are conveniently described by means of the effective-two-body formulation of the three-body theory \cite{ags67}. There, the amplitudes for the various two-fragment processes satisfy multichannel Lippmann-Schwinger-type equations
\begin{eqnarray}
{\cal T}_{\beta n, \alpha m} &=& {\cal V}_{\beta n, \alpha m} + \sum_{\nu, r}\; 
{\cal V}_{\beta n, \nu r}\; {\cal G}_{0;\nu r}\; {\cal T}_{\nu r, \alpha m} \label{efft} \\ 
{\cal V}_{\beta n, \alpha m} &=& \bar \delta_{\beta \alpha} \langle \chi_{\beta n} \mid G_0 \mid \chi_{\alpha m} \rangle + \sum_{\nu \neq \beta, \alpha} \langle \chi_{\beta n} \mid G_{0} T_{\nu}' G_{0} \mid \chi_{\alpha m} \rangle + \cdots,  \label{effp}
\end{eqnarray}
with $\bar \delta_{\beta \alpha} = 1- \delta_{\beta \alpha}$. The `form factor' $\mid \chi_{\alpha m} \rangle$ is a certain off-shell extrapolation of the bound state wave function $\mid \psi_{\alpha m} \rangle$. Already the first-order terms (`triangle amplitudes') in the expansion of ${\cal V}$ contain matrix elements of an operator $T_{\nu}'$, the most awkward part of which is the T-matrix $T_{\nu}^C$ describing Coulomb rescattering of the projectile off each of the target particles. Due to the singular nature of the latter, calculation of such terms is very difficult (this is particularly true if the rescattering particles have charges of opposite sign, because of the infinity of bound states). Indeed, the lack of applications of Faddeev-type approaches to atomic reactions is generally attributed to this fact \cite{de94}. 

Hence, the so-called Coulomb-Born Approximation (CBA) is usually made which consists in approximating $T_{\nu}^C$ by the Coulomb potential $V_{\nu}^C, \, \nu = \alpha, \beta, \gamma$. Recently, however, we have succeeded in calculating exactly all (on-shell) triangle amplitudes with the full (attractive and repulsive) Coulomb T-matrix, for all energies and scattering angles, for arbitrary bound state wave functions. In this way we could show that the CBA almost always fails badly for atomic triangle amplitudes \cite{akmr95,akm96,akm97}. (I mention that we have also derived (semi-)analytical approximations for these triangle amplitudes, valid for the higher energies considered here, which greatly simplify practical calculations but are much more accurate than the CBA \cite{akm97,akm96b}). 

The exact triangle amplitudes have been employed in an investigation of direct and exchange scattering of ${\rm H}^+$ off hydrogen atoms. For this purpose we have written Eqs. (\ref{efft}) in K-matrix approximation. When the resulting two-dimensional integral equations are transformed to the impact parameter representation they lead to a system of coupled algebraic equations \cite{a94}. The K-matrix has been approximated by the effective potential (\ref{effp}) were we keep the $0^{\rm th}$ and the full $1^{\rm st}$ order terms. For the latter, numerically calculated quantities the transformation to the impact parameter representation has been performed by quadrature. Our results provide a simultaneous, satisfactory description of differential cross sections for both elastic scattering and electron exchange; likewise the total exchange cross section is well-reproduced \cite{akm98}.

\subsection{Proton-Deuteron Scattering in the Integral Equations Approach}

The motivation behind 30 years of investigation of nucleon-deuteron ($Nd$) scattering rests on the hope that a detailed understanding may lead to deeper insights into the properties of $2N$ and $3N$ forces etc. Experimentally the proton-induced reaction ($pd$) is preferred; but reliable and sophisticated calculations have become available only recently, for energies below the deuteron breakup threshold \cite{bsz90,kvr96}. Some of the underlying reasons are pointed out in \cite{ams98}. Of course, the common practice of comparing $nd$ calculations with $pd$ data, though {\em in praxi} rather successful, is clearly unsatisfactory from principle point of view.

We \cite{ams98} have obtained the first reliable results for $pd$ scattering above the deuteron breakup threshold, for the Paris potential. We used the screening and renormalisation approach \cite{as96}. Comparison with elastic scattering cross section and polarisation data shows a reasonably good absolute agreement; but a much better reproduction is achieved of the difference between experimental $pd$ and $nd$ observables where available. Further progress requires, in particular, excellent, low-rank, separable representations of more modern potentials.

\begin{acknowledge}
Most of the results presented here have been worked out in collaboration with G.V. Avakov, B.F. Irgaziev, A.S. Kadyrov, A. M. Mu\-kha\-med\-zha\-nov, and A. I. Sattarov. 
\end{acknowledge}

\end{document}